# Origin of the type-II Weyl state in topological antiferromagnetic $YbMnBi_2$

Xiao-Sheng Ni,[1] Cui-Qun Chen,[1] Dao-Xin Yao,[1,*] and Yusheng Hou [1,†]

[1]State Key Laboratory of Optoelectronic Materials and Technologies, Center for Neutron Science and Technology, School of Physics, Sun Yat-Sen University, Guangzhou, 510275, China

## Abstract

Recently, there are heavy debates concerning the topological nature of $YbMnBi_2$ being a type-II Weyl semimetal with Mn moments having a canting angle of $10^o$ or a Dirac semimetal without canting. In this paper, we systematically study the magnetic and topological properties of $YbMnBi_2$ using density functional theory calculations. Our results unveil that bulk $YbMnBi_2$ is a Dirac semimetal with a collinear antiferromagnetic spin order. By canting the magnetic moments of Mn ions, we surprisingly find that $YbMnBi_2$ enters into the type-II Weyl state with a canting angle of $2^o$. As Bi vacancies in Mn-Bi-Mn bonds can produce sizable Dzyaloshinskii-Moriya interactions and thus cant the magnetic moments of Mn ions, we argue that the possible Bi vacancies induce a topological phase transition from a Dirac semimetal to a type-II Weyl semimetal in the realistic $YbMnBi_2$. Our work provides an insight into the realizations of the type-II Weyl state in antiferromagnetic topological materials.

Email: houysh@mail.sysu.edu.cn; yaodaox@mail.sysu.edu.cn

*Introduction.-* Topological materials, featuring topologically protected electronic states, have attracted tremendous attentions recently [1,2]. Among these materials, either Dirac semimetals (DSMs) or Weyl semimetals (WSMs) are of great interest and play important roles in leading to nontrivial quantum states for their special quasiparticles corresponding to high-energy particles [3]. Particularly, WSMs can be considered to originate from DSMs and exhibit paired nodes with opposite chiralities [3-6]. Because WSMs possess many peculiar properties, such as a small cyclotron effective mass, a low-energy linear dispersion depending on the electronic density as their square root [7], a high bulk carrier mobility [8], a non-saturating large magnetoresistance [9] and a nonlocal transport [10], they have promising applications in spintronics. After the conception of WSMs is put forward in 2011 [11] and the first WSM TaAs is discovered in experiments [12,13], a lot of efforts have been increasingly devoted to find new WSMs theoretically and experimentally [5]. Depending on whether or not they have electron and hole pockets, WSMs are further separated into type-I and type-II [4]. Although there are many researches focusing on the type-I WSMs such as the widely studied non-magnetic TaAs, the studies of type-II WSMs is still obscure due to the lack of candidates.

Excitingly, the recently synthesized antiferromagnet $YbMnBi_2$ provide a new candidate of Type-II WSMs [14] that is distinctly different from the previously studied Te-based compounds [3,15,16]. However, there has been a controversy in recent literatures about the topological nature of $YbMnBi_2$, which is associated with the canting of the magnetic moments of Mn ions. By comparing the band structures and basic properties of $YbMnBi_2$ and a structurally similar DSM candidate, $EuMnBi_2$ [17], S. Borisenko *et al.* firstly reported that $YbMnBi_2$ is a type-II WSM and its magnetic moments have a canting angle of 10º away from the *c* axis [14]. M. Chinotti *et al.* also confirmed the type-II WSM state of $YbMnBi_2$ after performing an optical investigation [18]. Later, J. Y. Liu *et al.* affirmed the type-II Weyl state in $YbMnBi_2$ by investigating the interlayer transport properties [19]. However, A. Wang *et al.* reported that the magnetic moments of Mn ions oriented along the *c* axis [20]. Besides, D. Chaudhuri *et al.* tended to believe that the optical experiment dates in Ref. [18] could be explained without canting the magnetic moments of Mn ions [21]. After studying the magnetic structure of YbMnBi$_2$, J. R. Soh *et al*. ruled out the existence of Weyl fermions and claimed YbMnBi$_2$ as a DSM [22]. These debates indicate the vital role of the canting of the magnetic moments of Mn ions in determining the topological nature of YbMnBi$_2$. Therefore, a comprehensive study on the magnetic properties of YbMnBi$_2$ is urgently needed to make clear its topological nature.

In this paper, we employ density functional theory (DFT) calculations to systematically investigate the magnetic and topological properties of YbMnBi$_2$. Our results demonstrate that bulk YbMnBi$_2$ is a DSM with a collinear antiferromagnetic (AFM) spin order. However, our further calculations reveal that YbMnBi$_2$ is a type-II WSM when the magnetic moments of Mn ions have a canting angle of $10^o$ away from the *c* axis, consistent with the experimental observation in Ref. [14]. Inspired by this, we cant the magnetic moments of Mn ions from $0^o$ to $16^o$ and find that YbMnBi$_2$ has already entered into the type-II Weyl state with a canting angle of $2^o$. According to our DFT calculated sizable Dzyaloshinskii-Moriya (DM) interactions in the Mn-Bi-Mn bonds with Bi vacancies and the fact that DM interactions can cant the magnetic moments in antiferromagnets, we argue that Bi vacancies in experiments can tune the topological nature of the realistic YbMnBi$_2$ from a DSM to a type-II WSM. Our work unravels the underlying mechanism for the type-II Weyl state in the realistic YbMnBi$_2$, providing a new perspective to study the Weyl state in AFM topological materials.

*Computational details.-* Our DFT calculations are performed by using Vienna *ab initio* simulation package (VASP) [23] with the generalized gradient approximation (GGA) in the framework of the Perdew-Burke-Ernzerhof (PBE) functional [24]. The projector-augmented wave pseudopotentials are adopted to describe the core-valence interaction. We choose an energy cutoff of 550 eV and a k-mesh grid of 12×12×5 centered at the gamma point. In structural relaxations, we use experimentally measured lattice constants and relax atomic positions with a force convergence criterion of 0.01 eV/Å. we utilize *U*=4 eV and $J_H$=1 eV to describe the strong correlation effects among $d$ electrons of Mn [25-27]. Using these parameters, we obtain a magnetic moment of 4.34 $\mu_B$/Mn for Mn ions, which shows a good agreement with the experimental value (see Fig. S1 in Supplemental Materials (SM) [28]). Spin-orbital coupling (SOC) is included in calculating band structures. With the DFT calculated band structures being well reproduced by wannier90 (see Fig. S2 in SM [28]), we employ wannier90 [29] and wanniertools [30] to analyze the topological properties of YbMnBi$_2$.

*Results.*- Bulk YbMnBi$_2$ has the tetragonal P4/nmm (No.129) group space [14] and its lattice constants measured by neutron diffraction at 4 K are a=b=4.46 Å and c=10.73 Å [20]. Below the Neel temperature $T_N$=290 K, its magnetic ordering is a C-type AFM spin order [22]. Moreover, its magnetic moment measured from experiments is 4.3 $\mu_B$/Mn [20], slightly smaller than SrMnBi$_2$ and CaMnBi$_2$ [31]. Fig. 1a depicts the crystal structure and AFM spin order of bulk YbMnBi$_2$. Note that there are two inequivalent Bi sites within the consideration of space group symmetry. For the convenience in the following discussions, we denote them as Bi1 and Bi2, respectively.

*Table I. Here are the DFT calculated Heisenberg exchange interaction parameters, NN DM interaction vector and single-ion anisotropy constant K of Eq. (1). Parameters $S^2J$, $S^2D$ and $S^2K$ are all in units of meV.*

| $S^2J_1$ | $S^2J_2$ | $S^2J_3$ | $S^2J_4$ | $S^2J_5$ | $S^2J_R$ | $S^2K$ | $S^2D_x$ | $S^2D_y$ | $S^2D_z$ |
|---|---|---|---|---|---|---|---|---|---|
| 49.03 | 10.32 | -6.74 | -4.2 | -2.15 | -0.54 | 2.20 | 0.08 | -0.55 | -0.01 |

To effectively capture the magnetic ground state properties of bulk YbMnBi$_2$, we use a spin Hamiltonian consisting of Heisenberg exchange interactions, DM interactions and the single-ion anisotropy. This spin Hamiltonian is written as follows:

$$H_{spin} = \sum_{ij} J_{ij} \mathbf{S}_i \cdot \mathbf{S}_j + \sum_{ij} \mathbf{D}_{ij} \cdot (\mathbf{S}_i \times \mathbf{S}_j) - K \sum_i S_{iz}^2 \quad (1).$$

In Eq. (1), $J_{ij}$ is the Heisenberg exchange interactions parameter; $\mathbf{D}_{ij}$ is the DM interaction vector and $K$ represents the single-ion anisotropy constant. We adopt the least-squares fitting technique [32] to calculate Heisenberg exchange parameters $J$ and the four-state method [33,34] to obtain the DM interaction vector $\mathbf{D}$. The DFT calculated Heisenberg exchange parameters $J$ are listed in Table I. One can see that the nearest neighbor (NN) $J_1$ is AFM and dominants over other long-range Heisenberg exchange interactions. The NN DM interaction is exactly vanishing due to the presence of inversion symmetry. Because there is a lack of inversion symmetry between the second-NN Mn pairs, the second-NN DM interactions can be nonzero in principles. Our calculations indicate that the second-NN DM interactions are really not zero, however very weak compared with the Heisenberg exchanges $J_1$ and $J_2$ (see Table I). Finally, the calculated single ion anisotropy constant is $S^2K = 2.20$ meV. This indicates that YbMnBi$_2$ has an out-of-plane magnetic easy axis. Overall, the magnetic interactions of bulk YbMnBi$_2$ can be characterized by the dominating Heisenberg exchange interactions together with a non-negligible single-ion anisotropy.

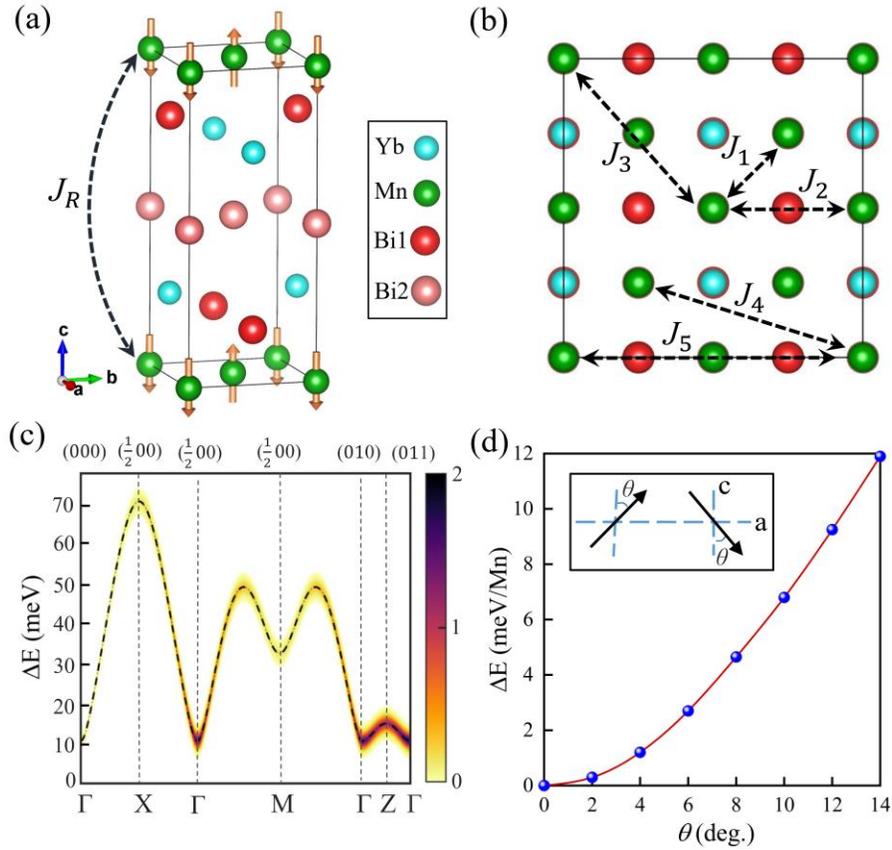

*FIG. 1. (a) The crystal and magnetic structures of bulk YbMnBi$_2$. Blue, green and red spheres represent Yb, Mn and Bi atoms, respectively. Magnetic moments of Mn ions are depicted by the golden arrows. The interlayer exchange path of $J_R$ is shown by the dashed black line. (b) The ab-plane exchange paths of $J_i$ (i=1,2,3,4 and 5) are indicated by the dashed black lines. (c) Spin-wave dispersion obtained by spinW [35] with the magnetic interaction parameters listed in Table I. (d) The dependence of the DFT calculated total energy on the canting angle, θ. The case of θ=0 is set as the reference. The inset shows the definition of θ.*

Based on the DFT calculated magnetic interaction parameters (Table I), we study the spin dynamics of Hamiltonian [Eq. (1)] under the linear spin wave approximation [35]. Our result shows that bulk YbMnBi$_2$ has a magnetic ground state of the collinear C-type AFM spin order (Fig. 1a), consistent with the DFT calculations. Fig. 1c depicts the corresponding spin-wave dispersion. We note that there is a small difference between our obtained spin-wave dispersion and the experimentally measured one in Ref. [22]. The huge peak centering at X distinctly reflects the strong intralayer AFM Heisenberg exchange interactions, while the small peak located at Z is consistent with the weak interlayer Heisenberg exchange interaction. Especially, the spin-wave gap of about 10 meV is well consistent with the experimentally measured one [36]. In short, our spin-wave analysis indicates the appropriate choice of the on-site Coulomb correlation $U$ in our DFT calculations to capture the key aspects of magnetic properties in YbMnBi$_2$.

Since the canting of the magnetic moments of Mn ions is the spotlight in the controversy of the topological nature of YbMnBi$_2$ [20,21,37,38], we wonder if the canting can appear in bulk YbMnBi$_2$. To this end, we compare the DFT calculated total energies of various cases where the magnetic moments of Mn ions have different canting angles. As shown in the inset of Fig. 1d, the canting angle, θ, is defined as tilting the magnetic moments of Mn ions from the *c* axis to the *a* axis. Fig. 1d clearly shows that the DFT calculated total energy increases monotonously with θ. This indicates that the most stable alignment of the magnetic moments of Mn ions is along c axis. That is to say, the magnetic moments of ions have no canting in bulk YbMnBi$_2$.

Here, we study the band structure of bulk YbMnBi$_2$ in order to gain insights into its topological nature. As shown in Fig. 2a-2c, there are several Dirac cones with a linear dispersion near the Fermi level. To examine their sources in the real space, we project the wave functions of the bands to Yb, Mn and Bi atoms, respectively. Fig. 2a-2b demonstrate the following important features in the energy window (from -2.0 to 2.0 eV) of interest: (I) there is fairly small contribution from Yb atoms to the bands; (II) a large portion of electronic states near the Fermi level is mainly from Mn atoms; (III) Bi2 atoms (the light red spheres in Fig. 1a) contribute most to the bands for the Dirac cones while Bi1 atoms have almost no contribution. These imply that Bi2 atoms may play an important role in determining the topological nature of YbMnBi$_2$.

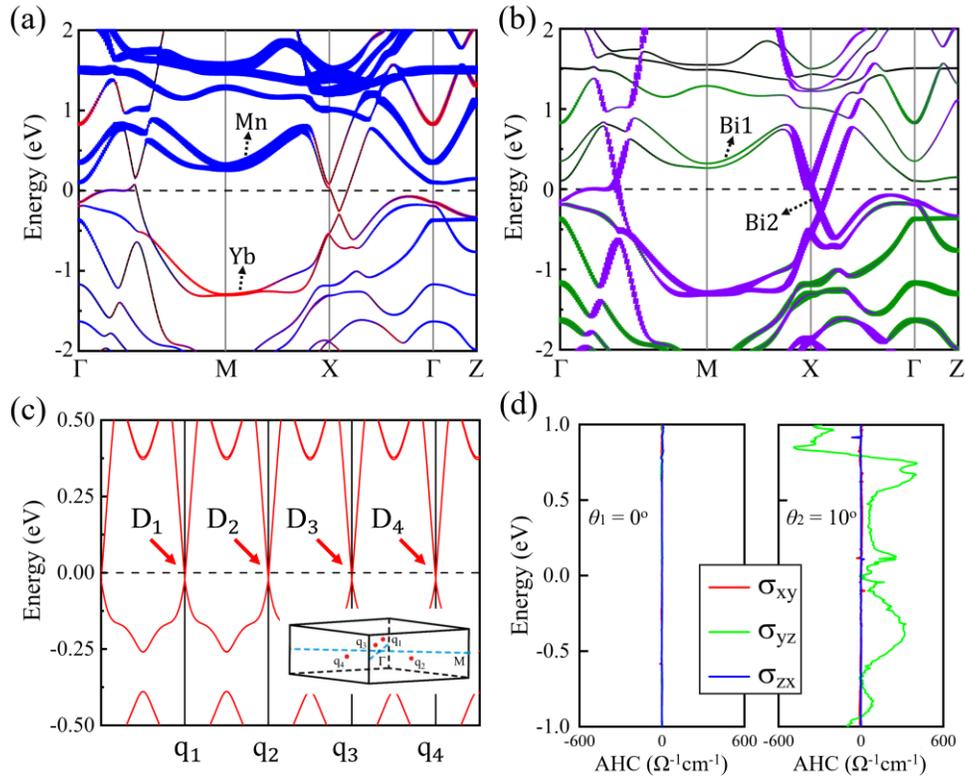

*FIG. 2. (a)-(b) Band structures projected to Yb (red), Mn (blue), Bi1 (green), Bi2 (violet) atoms. The width of the colorful lines indicates the weight of the contribution from different atoms. (c) The four Dirac nodes $D_i$ (i=1,2,3 and 4) near the Fermi Level in the first Brillouin zone of YbMnBi$_2$ without canting magnetic moments of Mn ions. Here, $q_1$=(-0.272, 0.271, 0.123), $q_2$=(0.271, 0.272, -0.122), $q_3$=(0.272, -0.272, 0.123) and $q_4$=(-0.272, -0.272, -0.124) are the coordinates of $D_1$, $D_2$, $D_3$, and $D_4$ in the reciprocal space, respectively. The inset shows the positions of $q_1$ to $q_4$ in the first Brillouin zone. (d) The AHCs of the canting angles $\theta_1$=0° and $\theta_2$=10°. The individual components of AHC are shown by the different kinds of colors in the inset. The Fermi level is set to zero in (a)-(d).*

Considering that the magnetic moments of Mn ions are reported to have a canting angle of 0° (i.e., a collinear C-type AFM spin order) [22] or 10° [14] in different experiments, we explore the topological properties of bulk YbMnBi$_2$ with these two different canting angles by determining chirality. As shown in Fig. 2c, there are four Dirac nodes without chirality in YbMnBi$_2$ when the canting angle is zero. Correspondingly, its anomalous Hall conductivity (AHC) is vanishing (see the left panel in Fig. 2d). Hence, it is clearly shown that YbMnBi$_2$ is a DSM when it has the collinear C-type AFM spin order. However, when the magnetic moments of Mn ions have a canting angle of 10°, we find that YbMnBi$_2$ has four pairs of type-II Weyl nodes with nonzero charities at the Fermi level (see Fig. 3g and Table SI in SM [28]). Accordingly, a nonvanishing AHC clearly shows up when the Fermi level is located in the vicinity of the type-II Weyl nodes (see the right panel in Fig. 2d). Therefore, it comes to the conclusion that YbMnBi$_2$ is a type-II WSM when the magnetic moments of Mn ions have a canting angle of 10°, consistent with the experimental observation in Ref. [14].

*Discussion.-* When the magnetic moments of Mn ions in YbMnBi$_2$ have a canting angle of 0°, the bands are doubly degenerated at least, due to the collinear C-type AFM spin order. Such doublet degeneracy of the bands is kept even when SOC is taken into account (see Fig. S3 in SM [28]). Therefore, YbMnBi$_2$ could be a DSM rather than a WSM. Our calculations show that it is indeed a DSM in this case. By contrast, when the magnetic moments of Mn ions have a canting angle of 10º, a weak ferromagnetism is induced in the collinear C-type AFM spin order. Consequently, the doublet degeneracy originating from the collinear C-type AFM spin order is lifted by the induced weak ferromagnetism. Actually, we indeed see that the four-fold degenerate Dirac nodes change to the doubly degenerate type-II Weyl nodes with tilted cones (see Fig. 3g). Furthermore, Fig. S5-S7 in SM [28] clearly show that the Weyl nodes in the canting angles of 2º, 10º, and 16º are type-II. Thus, the net ferromagnetism from the canting of the magnetic moments of Mn ions can give rise to the type-II WSM state in YbMnBi$_2$ when the canting angle is greater than 2º.

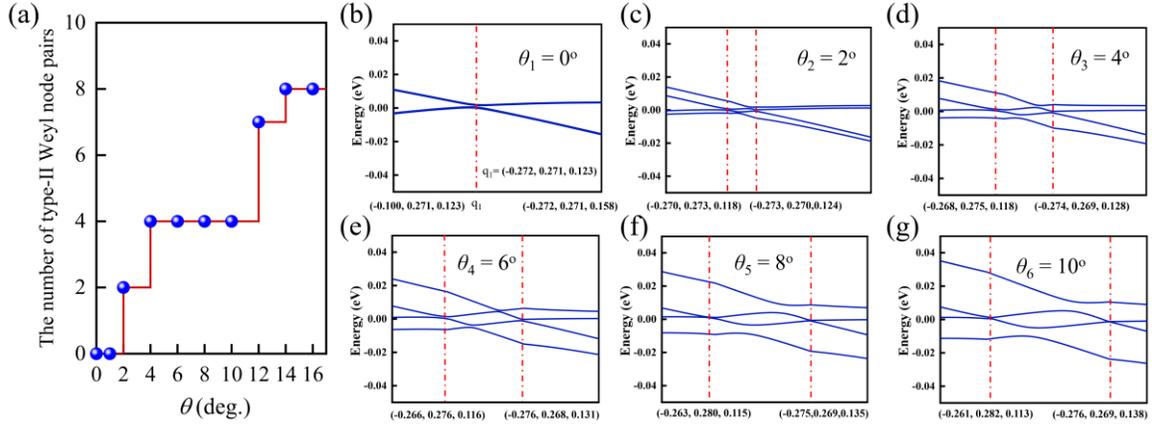

FIG. 3. (a) Dependence of the number of type-II Weyl node pairs on the canting angle ranging, $\theta$, from 0º to 16º in YbMnBi$_2$. (b)-(g) The evolution of band structures in the vicinity of the Dirac node or Weyl nodes pairs as the canting angle goes from 0º to 10º. The vertical dashed red lines indicate the positions of the type-II Weyl nodes and their coordinates in the reciprocal space are shown by the three numbers in the brackets.

To find out the relationship between the canting angle and the type-II Weyl state in YbMnBi$_2$, we investigate the dependence of the number of Weyl node pairs on the canting angle from 0º to 16º. From Fig. 3a, we can see that YbMnBi$_2$ has two pairs of the type-II Weyl nodes when the canting angle is just a small value of 2º. This shows that a topological phase transition is induced by a small canting of the magnetic moments of Mn ions in YbMnBi$_2$. Additionally, we see that more and more pairs of the type-II Weyl nodes appear when the canting angle increases. Interestingly, a similar mechanism is discussed in investigating the canting of magnetic moments to generate the type-II Weyl state in another topological antiferromagnet DyPtBi [39].

Now the crucial issue for the appearance of the type-II Weyl state in YbMnBi$_2$ is the underlying physical mechanism for which the magnetic moments of Mn ions cant. In fact, as experimentally observed in Ref. [40], there are various defects in YbMnBi$_2$ samples. These defects could be the factors that induce the canting of the magnetic moments of Mn ions. After investigating the possible defects of similar materials [41,42], we expect that Bi vacancies could exist in YbMnBi$_2$. As Bi1 atom is closer to the magnetic Mn layer than Bi2 atoms and its vacancies break the local inversion symmetry of the complex of Mn and Bi1 atoms, Bi1 vacancy could effectively introduce a strong DM interaction in YbMnBi$_2$. Considering that DM interactions can induce a weak ferromagnetism through canting magnetic moments in antiferromagnets [43,44], we study the effect of Bi1 vacancies on the magnetic and topological properties of YbMnBi$_2$. To this end, we use a 3×3×1 supercell and remove one Bi1 atom from this supercell (Fig. S4 in SM [28]). In this case, the corresponding defect density of Bi is 2.78% (see Part IV in SM [28]). As a direct influence of the Bi1 vacancies, structural relaxions reveal that Mn ions will move away from the inversion symmetry site (Fig. S4 in SM [28]). Once the DM interaction is obtained, we can estimate the canting angle, $\theta$, using the following expression [45-47]:

$$\tan\theta = |D_z/2J| \qquad (2).$$

In Eq. (2), $\theta$, $D_z$ and $J$ are the canting angle, the $z$-axis component of DM interaction and the Heisenberg exchange interaction, respectively.

We investigate the DM interactions of the NN Mn pair closest to the Bi1 vacancies. The NN DM interaction vector under such a circumstance is calculated to be $(D_x, D_y, D_z)$ = (0.70, -0.16, 2.99) meV. Using Eq. (2) and taking this DM interaction together with the NN Heisenberg exchange interaction $J_1$ = 49.03 meV, we obtain a canting angle $\theta$ = 1.74°. Note that the canting angle is zero in bulk YbMnBi$_2$ without defects, because the DM interaction between the NN Mn pair is forbidden by the inversion symmetry. Hence, Bi1 vacancy can result in the canting angle changing from 0º to 1.74º. This is highly consistent with the experimentally observed signals of magnetic moments canting in YbMnBi$_2$ [14,48]. Especially, 1.74º is quite close to 2º that leads to a topological phase transition from DSM to WSM in YbMnBi$_2$ (Fig. 3a). Thus, we argue that Bi1 vacancies could tilt the magnetic moments of Mn ions and cause the appearance of the type-II Weyl state in YbMnBi$_2$. It is worth noting that the key role of Bi atoms/vacancies in determining the topological properties of YbMnBi$_2$ is consistent with the studies in Ref. [49,50].

*Conclusions.-* In summary, we have investigated the magnetic properties and topological nature of YbMnBi$_2$ by means of a systematical DFT study. Our calculations unveil that bulk YbMnBi$_2$ is a DSM with a collinear C-type AFM spin order, consistent with the experimental observations. By canting the magnetic moments of Mn ions, we demonstrate that the type-II Weyl state shows up in YbMnBi$_2$ with a wide canting angle from 2º to 16º. This indicates that the canting of the magnetic moments of Mn ions induces a topological phase transition from DSM to WSM in YbMnBi$_2$. In addition, we find that Bi vacancies in Mn-Bi-Mn bonds can produce the sizable DM interactions. As DM interactions tilt the magnetic moments, we argue that Bi vacancies could tune the topological nature of YbMnBi$_2$ from DSM to the type-II WSM. Our findings provide a new dimension for realizing the type-II Weyl state in topological antiferromagnets and will stimulate more theoretical and experimental investigations in this field.

**ACKNOWLEDGMENTS**

This project is supported by NSFC-12104518, NKRDPC-2017YFA0206203, NKRDPC-2018YFA0306001, NSFC-92165204, NSFC-11974432, GBABRF-2019A1515011337, Shenzhen Institute for Quantum Science and Engineering (Grant No. SIQSE202102), Leading Talent Program of Guangdong Special Projects and the Startup Grant of Sun Yat-Sen University (No. 74130-18841290). We thank Zhi-Hui Luo for the helpful discussions.